\documentclass[preprint]{aastex}
\usepackage{url,lineno}
\linenumbers

\newcommand{\HI}{H\,{\small{\sc I}}}
\newcommand{\HII}{H\,{\small{\sc II}}}

\slugcomment{Not to appear in Nonlearned J., 45.}

\shorttitle{Collapsed Cores in Globular Clusters}
\shortauthors{Djorgovski et al.}

\begin{document}


\title{\textit{Fermi} LAT Observation of Diffuse Gamma-Rays 
Produced Through Interactions between Local Interstellar Matter 
and High Energy Cosmic Rays}


\author{
A.~A.~Abdo\altaffilmark{2,3}, 
M.~Ackermann\altaffilmark{4}, 
M.~Ajello\altaffilmark{4}, 
W.~B.~Atwood\altaffilmark{5}, 
M.~Axelsson\altaffilmark{6,7}, 
L.~Baldini\altaffilmark{8}, 
J.~Ballet\altaffilmark{9}, 
G.~Barbiellini\altaffilmark{10,11}, 
D.~Bastieri\altaffilmark{12,13}, 
B.~M.~Baughman\altaffilmark{14}
% ...
}
% ...
\altaffiltext{2}{Space Science Division, Naval Research Laboratory, Washington, DC 20375, USA}
\altaffiltext{3}{National Research Council Research Associate, National Academy of Sciences, Washington, DC 20001, USA}
\altaffiltext{4}{W. W. Hansen Experimental Physics Laboratory, Kavli Institute for Particle Astrophysics and Cosmology, Department of Physics and SLAC National Accelerator Laboratory, Stanford University, Stanford, CA 94305, USA}
\altaffiltext{5}{Santa Cruz Institute for Particle Physics, Department of Physics and Department of Astronomy and Astrophysics, University of California at Santa Cruz, Santa Cruz, CA 95064, USA}
\altaffiltext{6}{Department of Astronomy, Stockholm University, SE-106 91 Stockholm, Sweden}
\altaffiltext{7}{The Oskar Klein Centre for Cosmo Particle Physics, AlbaNova, SE-106 91 Stockholm, Sweden}
\altaffiltext{8}{Istituto Nazionale di Fisica Nucleare, Sezione di Pisa, I-56127 Pisa, Italy}
\altaffiltext{9}{Laboratoire AIM, CEA-IRFU/CNRS/Universit\'e Paris Diderot, Service d'Astrophysique, CEA Saclay, 91191 Gif sur Yvette, France}
\altaffiltext{10}{Istituto Nazionale di Fisica Nucleare, Sezione di Trieste, I-34127 Trieste, Italy}
\altaffiltext{11}{Dipartimento di Fisica, Universit\`a di Trieste, I-34127 Trieste, Italy}
\altaffiltext{12}{Istituto Nazionale di Fisica Nucleare, Sezione di Padova, I-35131 Padova, Italy}
\altaffiltext{13}{Dipartimento di Fisica ``G. Galilei", Universit\`a di Padova, I-35131 Padova, Italy}
\altaffiltext{14}{Department of Physics, Center for Cosmology and Astro-Particle Physics, The Ohio State University, Columbus, OH 43210, USA}
% ...

%%\author{S. Djorgovski\altaffilmark{1,2,3} and Ivan R. King\altaffilmark{1}}
%%\affil{Astronomy Department, University of California, Berkeley, CA 94720}


%%\altaffiltext{1}{Visiting Astronomer, Cerro Tololo Inter-American Observatory. 
%% CTIO is operated by AURA, Inc.\ under contract to the National Science Foundation.}
%%\altaffiltext{2}{Society of Fellows, Harvard University.}
%%\altaffiltext{3}{present address: Center for Astrophysics, 60 Garden Street, Cambridge, MA 02138}


\begin{abstract}
Observations by the Large Area Telescope (LAT)
on the \textit{Fermi} mission of diffuse $\gamma$-rays in a mid-latitude region 
in the third quadrant (Galactic longitude $l$ from $200\arcdeg$
to $260\arcdeg$ and latitude $\left| b \right|$
from $22\arcdeg$ to $60\arcdeg$)  
are reported. The region contains no known large molecular cloud 
and most of the atomic hydrogen is within 1~kpc of the solar system. 
The contributions of $\gamma$-ray point sources 
and inverse Compton scattering are estimated and subtracted. 
The residual $\gamma$-ray intensity exhibits 
a linear correlation with the atomic gas column density 
in energy from 100~MeV to 10~GeV. 
The measured integrated $\gamma$-ray emissivity is 
$(1.63 \pm 0.05) \times 10^{-26}~{\rm photons~s^{-1}~sr^{-1}~H\mathchar`-atom^{-1}}$ 
and 
$(0.66 \pm 0.02) \times 10^{-26}~{\rm photons~s^{-1}~sr^{-1}~H\mathchar`-atom^{-1}}$
above 100~MeV and above 300~MeV, respectively, 
with additional systematic error of $\sim 10$~\%.
The differential emissivity from 100~MeV to 10~GeV
agrees with calculations based on
cosmic ray spectra consistent with those directly measured,
at the 10~\% level.
The results obtained indicate that cosmic ray nuclei spectra 
within 1~kpc from the solar system in regions studied are 
close to the local interstellar spectra inferred from 
direct measurements at the Earth within
$\sim 10~\%$.

\end{abstract}





\section{Introduction}

The diffuse high energy $\gamma$-ray emission ($E \ge 30~{\rm MeV}$)
has been interpreted to be a superposition of $\gamma$-rays 
produced via interactions between cosmic rays (CRs) 
and interstellar matter, inverse Compton (IC) 
scattering of interstellar soft photons off CR electrons, 
and the extragalactic diffuse $\gamma$-ray emission. 
The first component, if distinguished from the others, 
will enable using high-energy $\gamma$-ray observations 
for the study of the distribution of CRs and the interstellar medium. 
The distribution of neutral atomic hydrogen (\HI)
is traced by 21~cm line surveys 
and the molecular hydrogen distribution is derived 
indirectly using 2.6~mm line observations 
of carbon monoxide (CO).
The total gas column density
can also be traced indirectly from extinction and reddening
by dust.
Thus the spectrum and the flux of CRs 
can be obtained from sufficiently sensitive observations 
of high energy $\gamma$-rays. 
The observation of diffuse $\gamma$-rays away 
from the Galactic plane (Galactic latitude $\left| b \right| \ge 10\arcdeg$) 
is suitable for studying local CRs, 
since diffuse $\gamma$-rays in such regions are
less affected by contamination from strong point sources, 
and most of the gas along the line of sight is local.
The SAS-2 \citep[e.g.,][]{Fichtel1978}
and COS-B observations \citep[e.g.,][]{Lebrun1982}
indicated a correlation between the $\gamma$-ray intensities 
and the total gas column densities at medium Galactic latitudes. 
\citet{Sreekumar1998} and \citet{Strong2004}
showed a good correlation between the $\gamma$-ray intensities 
and model calculations in their analyses of the extragalactic diffuse emission 
observed by EGRET onboard the 
\textit{Compton Gamma-Ray Observatory}. 
Despite these early studies, the flux and spectrum of local CRs
deduced from $\gamma$-ray intensity remain uncertain, 
due to the possible contamination from unresolved point sources
and 
uncertainties in modeling the IC contribution because of the
large scale height of CR electrons and the reprocessing
of the interstellar radiation by dust.
Although CR nuclei in the vicinity of the solar system 
are thought to have spectral distributions and intensities
similar to those measured at the Earth as reported by
a number of $\gamma$-ray observations
\citep[e.g,][]{Hunter1997,Digel2001a},
data above 1~GeV,
which are crucial to distinguish
CR nuclei spectra from that of CR electrons,
have not been good enough due to
the limited photon statistics and relatively limited energy coverage
of these early missions.

The situation has improved significantly 
with the recent launch of the \textit{Fermi} Gamma-ray Space Telescope 
on 2008 June 11. The \textit{Fermi} LAT (Large Area Telescope) 
has a sensitivity that is more than an order of magnitude 
better than that of EGRET
and enables resolving point sources 
and studying the diffuse $\gamma$-rays with unprecedented sensitivity. 
Recent advances of a CR propagation code GALPROP 
\citep[e.g.,][]{Strong1998},
which had been developed through comparisons 
with the EGRET data,
allow us to predict and subtract IC emission 
and correlate $\gamma$-ray emission with interstellar matter
more accurately. 

In this paper, we present \textit{Fermi} LAT observations 
of diffuse $\gamma$-rays in a mid-Galactic latitude region 
in the third quadrant (Galactic longitude $l$ from $200\arcdeg$ to 
$260\arcdeg$ and $\left| b \right|$ from $22\arcdeg$ to $60\arcdeg$). 
As discussed in the following sections, 
most of the gas along the line of sight is local, 
nearby on the scale of the Milky Way. 
The contribution from IC emission is 
only about 10~\% of the total diffuse emission
and the LAT has already resolved 
five times as many $\gamma$-ray point sources 
as previous missions in this region. 
These facts enable us to evaluate the local CR flux 
and the spectrum with small systematic uncertainty.

\section{Observation and Data Reduction}

The LAT is the main instrument 
of the \textit{Fermi} Gamma-ray Space Telescope. 
It consists of $4 \times 4$ modules (towers) 
built with tungsten foils and silicon microstrip detectors 
to measure the arrival directions of 
incoming $\gamma$-rays, and a hodoscopic cesium iodide calorimeter 
to determine the photon energies. 
They are surrounded by 89 segmented plastic scintillators 
serving as an anticoincidence detector 
to reject charged particle events. 
Details of the LAT instrument 
and pre-launch expectations of the performance
can be found in 
\citet{Atwood2009}.
The excellent sensitivity of the LAT is 
exemplified by initial publications such as \citet{Abdo2008}.

Routine science operations with the LAT 
began on 2008 August 4.  
We have accumulated events from 
2008 August 4 to 2009 January 31 to study diffuse $\gamma$-rays. 
During this time interval the LAT was operated 
in sky survey mode nearly all of the time; 
in this observing mode the LAT scans the sky, 
obtaining complete sky coverage every two orbits and 
relatively uniform exposures over time.  
We used the standard LAT analysis software, 
{\bf ScienceTools}
\footnote{
available from the Fermi Science Support Center
(\url{http://fermi.gsfc.nasa.gov/ssc/})}
version {\bf v9r11}, 
and applied the following event selection criteria:  
(1) events have the highest probability of being photons, 
i.e., they are categorized as so-called diffuse class
\citep{Atwood2009},
(2) the reconstructed zenith angles of the arrival direction 
of photons are selected to be less than $105\arcdeg$, 
in order to exclude periods where the Earth
enters the LAT field of view, and
(3) the center of the LAT field of view is within 
$39\arcdeg$ from the zenith in order 
not to include the data taken in the pointed observation mode,
because it has increased contamination from Earth albedo $\gamma$-rays. 
We also eliminated the period of time during which 
the LAT detected two bright GeV-emitting GRBs,
i.e., GRB080916C \citep{Abdo2009a} and GRB081024B
\citep{Omodei2008}.
We then generated count maps (using {\bf gtbin} in {\bf ScienceTools}) 
and exposure maps (using {\bf gtexpcube}) 
in 13 logarithmically-sliced energy bins 
from 100~MeV to 9.05~GeV.
A post-launch response function {\bf P6\_V3\_DIFFUSE},
which was developed to account for the $\gamma$-ray
detection inefficiencies that are correlated with trigger rate,
was used in exposure calculations. 
These count and exposure maps were prepared in Cartesian coordinates
in $0.5\arcdeg \times 0.5\arcdeg$ binning,
and then transformed into HEALPix 
\footnote{\url{http://healpix.jpl.nasa.gov}}
\citep{HealPix2005}
equal area sky maps of order=7.
They are used below to correlate the $\gamma$-ray 
intensities with the column densities of 
atomic gas along the line of sight;
the $\gamma$-ray intensity is calculated as the ratio 
of the counts and the exposures for each energy bin.

\section{Data Analysis}

\subsection{Subtraction of Inverse Compton and Point Sources}

To distinguish $\gamma$-rays produced in the interstellar medium 
from others, we referred to the GALPROP prediction 
of IC emission and an LAT source list for 6 month data
\footnote {internally available to the LAT team}.
This list was produced using a similar procedure used to obtain
the LAT Bright Source List described by
\citet{Abdo2009b}.
It covers the same period of time 
as that of our data set and contains 740 point sources 
with significance more than 5 $\sigma$.  
We adopted positions and spectral parameters 
from this list (single power-law model in 100~MeV--100~GeV) 
to estimate and subtract the photons
from point sources to diffuse $\gamma$-ray emission. 
GALPROP \citep[e.g.,][]{Strong1998}
is a set of programs 
to solve the CR transport equation within our Galaxy 
and predict the $\gamma$-ray emission produced 
via interactions of CRs with interstellar matter 
(nucleon-nucleon interaction and electron bremsstrahlung) 
and soft photons (IC scattering). 
IC emission is calculated from the distribution of (propagated) 
electrons and the interstellar radiation fields 
developed by \citet{Moskalenko2006}.
Here we adopted the IC model map with version 
{\bf 54\_5gXvarh7S}
\footnote{
The GALPROP galdef ID of this version
is available at the website 
\url{http://galprop.stanford.edu}},
which was used in another \textit{Fermi} LAT
paper to study the diffuse $\gamma$-ray emission
in $10\arcdeg \le \left| b \right| \le 20\arcdeg$
\citep{Abdo2009d}.
The CR electron spectrum is adjusted to agree with
the directly-measured pre-Fermi spectrum in this GALPROP model.
In order to minimize the uncertainty of the contribution 
from IC emission on the diffuse $\gamma$-ray spectrum, 
we selected sky regions away from the Galactic center; 
the lower CR electron fluxes and interstellar radiation field 
will result in dimmer 
IC emission than that toward the Galactic center. 
We chose the third quadrant, 
Galactic longitude $l$ from $200\arcdeg$ to $260\arcdeg$ 
and the Galactic latitude $b$ from $-60\arcdeg$ to $-22\arcdeg$ 
and from $22\arcdeg$ to $60\arcdeg$. 
The region is free of known large molecular clouds;
Orion molecular clouds (Orion A and Orion B) and 
Monoceros molecular cloud complexes are located
in the region $l$ from $200\arcdeg$ 
to $220\arcdeg$ and $b$ from $-10\arcdeg$ to $-20\arcdeg$, 
and the Taurus/Perseus molecular clouds are in $l$ 
from $150\arcdeg$ to $185\arcdeg$ 
\citep[e.g.,][]{Dame2001,Digel1999,Digel2001b}.
Therefore the region described is suitable 
for correlating the $\gamma$-ray intensities 
with the local atomic-gas column densities.  

In Figure~1, we show $\gamma$-ray count maps 
above 100~MeV.
There are 52 sources in the LAT 6 month source list
in our region of interest,
more than five times as many sources
in the third EGRET catalog \citep{Hartman1999} 
in this region (nine sources). The diffuse $\gamma$-ray spectrum, 
after masking sources 
with circular regions of $1\arcdeg$ radius, 
is shown in Figure~2.
Atomic hydrogen column density maps of the same region
(see Section 3.2 for details) are given in Figure~3.
In Figure~2 and figures shown hereafter
(Figures~4--6),
the $\gamma$-ray intensities or CR fluxes multiplied by
$E^{2}$ (where $E$ is the 
center of each energy bin in logarithmic scale)
are presented.
Also presented in Figure~2 is
the contribution from IC emission 
predicted by GALPROP,
and the spillover from point sources 
outside the mask regions estimated 
(using {\bf gtmodel}) by the 
spectral parameters given in the source list.
Both the flux of estimated IC emission and the
residual point source contribution are less than 15~\% 
of the total diffuse emission above 100~MeV. 
We thus conclude that the uncertainty due to the IC 
and point source contributions is negligible 
after we subtract them from $\gamma$-ray data. 
Hereafter we analyze diffuse emission 
after masking point sources and subtracting IC emission and 
the residual contributions from point sources.

\subsection{Atomic Hydrogen Map}

Column densities $N$(\HI) of atomic hydrogen gas 
were calculated from existing radio surveys of the 
21~cm line of \HI.  We used the Leiden/Argentine/Bonn 
(LAB) Survey which merges the Leiden/Dwingeloo Survey 
\citep{Hartmann1997} 
with the Instituto Argentino de Radioastronomia Survey 
\citep{Arnal2000,Bajaja2005}
and covers the entire sky. 
We applied an optical depth correction 
under the assumption of a uniform spin temperature of 125~K 
and the cosmic microwave background intensity at 
1420~MHz of 2.66~K \citep[e.g.,][]{Hunter1994}.
The derived \HI\ column density maps 
of our region of interest are shown in Figure~3. 
Although major CO surveys such as the one by 
\citet{Dame2001} 
do not cover the region we analyzed, 
no large molecular cloud is known there 
and the molecular gas contribution is expected 
to be small due to the moderately high Galactic latitude. 
See the discussion by \citet{Dame2001}
for the completeness of their survey. 
Hereafter we assume that all the gas is in atomic form 
and traced by 21~cm radio surveys. 
Column densities of \HI\ in our region range from 
${\rm 1 \times 10^{20}~cm^{-2}}$ up to
${\rm 18 \times 10^{20}~cm^{-2}}$
and the optical depth correction is rather small;
the increase of the column densities from those 
for the optically thin case (infinite spin temperature)
is $\le 10$~\% in most directions.
On the assumption of a Galactic rotation curve by \citet{Clemens1985}
for the case of $R_{0} = 8.5~{\rm kpc}$ and
$\theta_{0} = 220~{\rm km~s^{-1}}$ 
(where $R_{0}$ and $\theta_{0}$ are the Galactocentric radius
and the orbital velocity of the local group of stars, respectively),
we infer that, 
in almost every direction in our region, 
more than 80~\% of the \HI\ 
along the line of sight is within 1~kpc of the solar circle.
Furthermore, by referring to the vertical density distribution of
\HI\ given by \citet{Dickey1990}, we can conclude that
more than 85~\% of atomic gas in the line of sight
is within 1~kpc of the solar system 
for $\left| b \right| \ge 22\arcdeg$.

\subsection{Correlation of $\gamma$-ray Intensities and Gas Column Densities }

The LAT point-spread function (PSF) 
strongly depends on the photon energy 
\citep[e.g.,][]{Atwood2009}
and the energy dependence of the angular size 
needs to be taken into account in data analysis. 
We convolved the map of \HI\ column densities 
obtained as described in Section 3.2 
using the GaDGET package \citep{Ackermann2008}
with the LAT PSF for 
each of our energy bins;
in the convolution we used the all sky map to
take account of the contribution
from outside the region for the analysis.
Since the typical angular size of the 
variation of column densities is a few degrees (see Figure~3), 
only maps for the lowest energy bands (less than a few hundred MeV) 
are noticeably smeared. 

The $\gamma$-ray intensities, after masking point sources 
with $1\arcdeg$ circular regions and subtracting the 
IC emission and the residual point source contributions,
are correlated with the \HI\ 
column densities in each energy band.
Both the $\gamma$-ray intensity map and the \HI\
column density map were prepared in
HEALPix equal area sky maps
of order=7, whose pixel size in solid angle is
$6.39 \times 10^{-5}$ steradian and is
close to that of ${\rm 0.5 \times 0.5~deg^{2}}$.
We found a linear relationship between 
$N$(\HI) and residual $\gamma$-ray intensities for
energies 
from 100~MeV to 10~GeV. 
Above 10~GeV the correlation is limited by photon statistics.
Figure~4 shows the correlation between
$\gamma$-ray intensities and the 
\HI\ column densities for four representative energy bands. 
The linear correlation
indicates that 
point source contributions are successfully subtracted 
and residual $\gamma$-rays mostly originate
from interstellar atomic gas 
through interactions with CRs, plus isotropic diffuse component 
(extragalactic diffuse $\gamma$-rays,
the residual particle background, and a possible residual
of IC emission).

By fitting the correlation in each energy band
with a linear function using a $\chi^{2}$ minimization,
we obtained the intensity of the isotropic diffuse component 
and the emissivity of atomic gas as the 
offset and the slope, respectively, 
as summarized in Table~1. 
Making the mask region larger, to $3\arcdeg$ radius, 
gives consistent fit parameters within statistical errors, 
confirming that the contribution of point sources is well modeled 
and subtracted. The obtained isotropic diffuse component 
(“offset” column in Table~1) agrees within 10--20~\%
with the "Isotropic" component given in \citet{Abdo2009d} 
which investigates
the medium-latitude diffuse emission.
We note that the isotropic diffuse components obtained here 
and in \citet{Abdo2009d}
include the residual background and thus
should be regarded as an upper limit of the 
true extragalactic diffuse $\gamma$-ray emission. 
We also note that the adopted IC model affects the 
spectral shape and the intensity of our isotropic component,
whereas it does not affect the emissivity significantly;
modifying the IC emission by $\pm 50$~\% changes the
offsets by 6~\%--7~\%, but alters
the slopes less than 3~\% except
the lowest two energy bands.
A detailed study of the extragalactic diffuse emission 
and the residual background using data for a larger sky area 
is underway and will be published elsewhere 
(A. A. Abdo et al. 2009, in preparation).
% \citep{Abdo2009e}.

So far, we have been neglecting the contribution from CR
interactions with ionized hydrogen (\HII).
The low-density ionized gas is unobservable, but can be
inferred from dispersion measures of pulsar signals 
in the radio band. According to the model of \citet{Cordes2002},
in the region we are studying, $N$(\HII) is only
${\rm (1\mbox{--}2) \times 10^{20}~cm^{-2}}$ and fairly smooth.
We thus conclude that the contribution
from ionized gas does not affect the obtained emissivity
significantly.

\section{Discussion}

With the approach described in Section~3, 
we succeeded in decoupling diffuse $\gamma$-rays 
related to the local atomic gas from point sources, 
the IC emission, and the isotropic diffuse component. 
The derived differential $\gamma$-ray emissivity from 
the local atomic gas is given in Figure~5. 
The systematic uncertainty of the effective area 
of the response we used ({\bf P6\_V3\_DIFFUSE})
is estimated
to be 10~\%, 5~\% and 20~\% at 100~MeV,
560~MeV, and 10~GeV, respectively,
and depend on the energy linearly in a logarithmic scale.
This systematic uncertainty is comparable with the statistical error,
and is indicated by the shaded area in the figure.
The integral emissivity above 100~MeV and 300~MeV is 
$(1.63 \pm 0.05) \times 10^{-26}~{\rm photons~s^{-1}~sr^{-1}~H\mathchar`-atom^{-1}}$ 
and
$(0.66 \pm 0.02) \times 10^{-26}~{\rm photons~s^{-1}~sr^{-1}~H\mathchar`-atom^{-1}}$, 
respectively, with an additional systematic uncertainty of $\sim 10~\%$.
These values can be compared with 
those reported by early measurements. 
SAS-2 \citep{Fichtel1978} obtained about 
$3 \times 10^{-26}~{\rm photons~s^{-1}~sr^{-1}~H\mathchar`-atom^{-1}}$
and COS-B \citep{Lebrun1982} reported 
$(1.67 \pm 0.24) \times 10^{-26}~{\rm photons~s^{-1}~sr^{-1}~H\mathchar`-atom^{-1}}$
above 100~MeV. 
The EGRET analysis of various directions 
toward large molecular clouds \citep{Digel2001a}
gives 
$(1.65\mbox{--}2.4)\times 10^{-26}~{\rm photons~s^{-1}~sr^{-1}~H\mathchar`-atom^{-1}}$ 
and
$(0.71\mbox{--}1.0)\times 10^{-26}~{\rm photons~s^{-1}~sr^{-1}~H\mathchar`-atom^{-1}}$
above 100~MeV and 300~MeV, respectively. 
While most of these early measurements
are consistent with the LAT data,
the emissivity obtained by the LAT
is much improved in photon statistics and energy range.

We can give constraints on the local CR spectrum 
by comparing the obtained emissivity 
with the model calculation of interactions between CRs 
and interstellar matter. 
Many evaluations of the $\gamma$-ray production due to CR interactions 
in the interstellar medium have been made, including 
\citet{Stecker1973,Stecker1989}, \citet{Dermer1986a,Dermer1986b}, 
\citet{Bertsch1993},
\citet{Mori1997}, \citet{Kamae2006} and \citet{Huang2007}.
In the calculation of neutral pion production and
decay $\gamma$-rays, most authors have computed the $\gamma$-ray flux
produced through interactions of
high-energy CR protons with proton targets.
The effects of heavy nuclei in both CRs and the target matter
are usually taken into account as a so-called nuclear enhancement factor
($\epsilon_{\rm M}$)
to multiply the proton-proton $\gamma$-ray yield.
Although the predicted $\gamma$-ray spectra
from proton-proton interactions 
assuming the same CR proton spectrum agree well
($\le 10~\%$)
among these works
\citep[e.g.,][]{Kamae2006}, the nuclear enhancement factors
differ by up to $\sim 30~\%$;
the factors range from 1.45 to 1.80--2.0 as compiled
by \citet{Mori2009}.
Among them, \citet{Dermer1986a,Dermer1986b} gives the lowest
$\epsilon_{\rm M}$ of 1.45, and \citet{Mori2009}
gives the highest $\epsilon_{\rm M}$ of 1.84 at the CR proton
kinetic energy of 10~GeV.
His higher value of $\epsilon_{\rm M}$ is attributed
to the adoption of recent CR spectral formulae by 
\citet{Honda2004} and the inclusion of nuclei heavier than
He in both the interstellar medium and the CR spectra.
We thus regard $\epsilon_{\rm M}$ by
\citet{Mori2009} as the most reliable.

In calculating the neutral pion production,
we used the
proton-proton interaction formalism 
by \citet{Kamae2006}.
They gave parameterized formulae of the
$\pi^{0}$ inclusive cross section and decay $\gamma$-ray
spectra for arbitrary proton kinetic energies
from 0.488~GeV to 512~TeV.
We adopted the proton local interstellar spectrum (LIS)
from the GALPROP model
with {\bf 54\_5gXvarh7S}
and calculated the $\gamma$-ray spectrum from nucleon-nucleon
interactions using formulae given by \citet{Kamae2006}
under the assumption
of the nuclear enhancement factor to be 1.84
as a representative value of those by
\citet{Mori2009}.
In order to calculate the electron bremsstrahlung,
we fully utilized GALPROP which
calculates the $\gamma$-ray spectrum
using a formalism by \citet{Koch1959} as explained by \citet{Strong2000}. 

The predicted emissivity at the solar sytem
(Galactocentric radius $R=8.5~{\rm kpc}$ and the height from the 
Galactic plane $z=0~{\rm kpc}$) is compared with our LAT measurement 
in Figure~5, 
and the LIS of proton,
electron and positron used in this model calculation 
is presented in Figure~6. 
Also shown is a compilation of some measurements 
of proton and electron spectra at the Earth. 
The proton and electron model spectra follow 
the observed ones above a few tens of GeV;
below this energy the solar wind lowers the observed fluxes. 
We note that the CR electron spectrum measured by the LAT
is somewhat harder than the GALPROP model
\citep{Abdo2009c}, but the effect on our analysis is negligible.
We also note that the bremsstrahlung at around 100~MeV
has comparable contributions from both primary electrons
and secondary electrons/positrons, as discussed by \citet{Porter2008}.
Their contributions are included in the computed spectrum 
shown in Figure~5.

The true LIS is somewhat uncertain due to solar activity.
To model this solar modulation effect
on the CR spectrum, 
the formula by
\citet{Gleeson1968} generally has been used,
in which a single parameter $\phi$ is introduced.
The proton LIS we adopted reproduces the observations at the Earth
with ${\rm \phi = 450~MV}$ as shown in Figure~6.
The same data can also be reproduced by a different 
formula of the LIS (dotted blue line) with ${\rm \phi = 600~MV}$
as described by \citet{Shikaze2007}.
We thus regard the difference between two models
as representing the uncertainty of the LIS;
they agree well ($\le 10~\%$) above 10~GeV and
differ by $\sim 20~\%$ at about 3~GeV. This affects the
calculated emissivity by $\sim 20~\%$ and
$\le 10~\%$ at about 100~MeV and above 1~GeV, respectively
\citep[e.g.,][]{Mori1997}.
We note that the LIS we adopted better reproduces 
the observed proton spectrum
above 20~GeV.
Although the true LIS below 1~GeV is highly uncertain,
these CRs do not contribute to the $\gamma$-ray emissivity 
above 100~MeV significantly.

As shown in Figure~5,
the emissivity measured by the LAT agrees
with the prediction from the assumed LIS
and the recent estimate
of $\epsilon_{\rm M}$
at the 10~\% level,
which is comparable with the statistical error and
the current systematic uncertainty 
of the LAT response. 
For reference, we also show
the $\gamma$-ray emissivity model obtained with
the lowest $\epsilon_{\rm M}$ 
\citep[among references in][]{Mori2009} of 1.45
that gives a predicted emissivity lower than
that observed in 100~MeV--10~GeV.
Since the nucleon-nucleon component is dominant in the emissivity
spectrum especially above 1~GeV,
the observed agreement between the LAT data and 
the model calculation (with the latest estimate of $\epsilon_{\rm M}$)
indicates that CR nuclei in the vicinity of the solar system
in regions observed
have spectral distributions and intensities close to those
of the LIS inferred from
measurements at the Earth within $\sim 10~\%$.
Although the constraint is rather weak,
the agreement down to 100~MeV also suggests that
highly uncertain low-energy (below a few hundred MeV)
CR electron and positron spectra are compatible with
our assumption (GALPROP model with {\bf 54\_5gXvarh7S})
shown in Figure~6. 

\section{Summary and Conclusions}

We report the observation of diffuse $\gamma$-rays 
in a mid-latitude region in the third quadrant 
using data from the first six months 
of \textit{Fermi} LAT science observations.
The region is away from the Galactic plane and the Galactic center, 
and contains no known large molecular cloud. 
Most of the atomic hydrogen is within 1~kpc of the solar system,
and thus the region is suitable for studying
the $\gamma$-ray emissivity of the local atomic gas and CR
spectra in the neighborhood of the solar system.
Thanks to the excellent performance of the LAT 
and recent developments of the CR propagation code 
and the interstellar radiation field model in GALPROP, 
we reliably estimated and subtracted the contribution 
from point sources and IC emission. 
The residual $\gamma$-ray intensities exhibit 
a linear relationship with the atomic gas column densities 
from 100~MeV to 10~GeV, indicating
that non-isotropic $\gamma$-rays are produced through interactions 
of CRs
with interstellar atomic gas.
The measurement of the emissivity of local atomic hydrogen 
has already surpassed those by past missions
in photon statistics and the energy range.  
It agrees with the prediction from
CR spectra assumed,
indicating that the CR nuclei spectra 
in the vicinity of the solar system 
in regions analyzed are close to the LIS
inferred from direct measurements at the Earth
within $\sim 10$~\%.
Low energy CR electron/positron spectra
are suggested to be compatible with our assumption.

\acknowledgments

The \textit{Fermi} LAT Collaboration acknowledges generous ongoing support 
from a number of agencies and institutes 
that have supported both the development 
and the operation of the LAT as well as scientific data analysis.  
These include the National Aeronautics and Space Administration 
and the Department of Energy in the United States, 
the Commissariat \`a l'Energie Atomique 
and the Centre National de la Recherche Scientifique / Institut 
National de Physique Nucl\'eaire et de Physique 
des Particules in France, the Agenzia Spaziale Italiana 
and the Istituto Nazionale di Fisica Nucleare in Italy, 
the Ministry of Education, Culture, Sports, Science and Technology (MEXT), 
High Energy Accelerator Research Organization (KEK) and 
Japan Aerospace Exploration Agency (JAXA) in Japan, 
and %%the K.~A.~Wallenberg Foundation, 
the Swedish Research Council and the Swedish National Space Board in Sweden.
Additional support for science analysis during the operations 
phase from the following agencies is also gratefully acknowledged: 
the Istituto Nazionale di Astrofisica in Italy 
and the K.~A.~Wallenberg Foundation in Sweden.

Some of the results in this paper have been derived using the
HEALPix \citep{HealPix2005} package.

\begin{figure}
\plotone{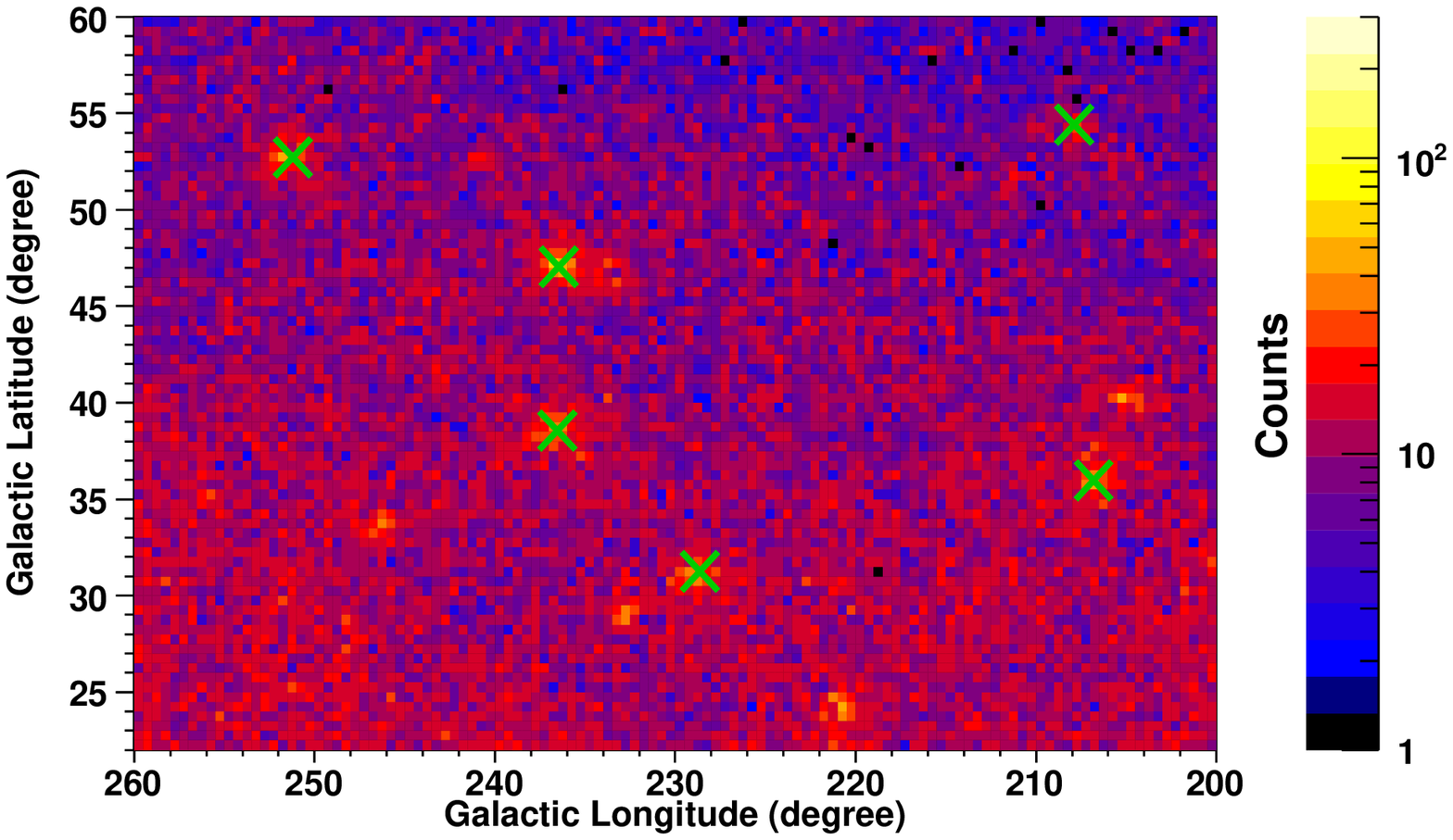}
\plotone{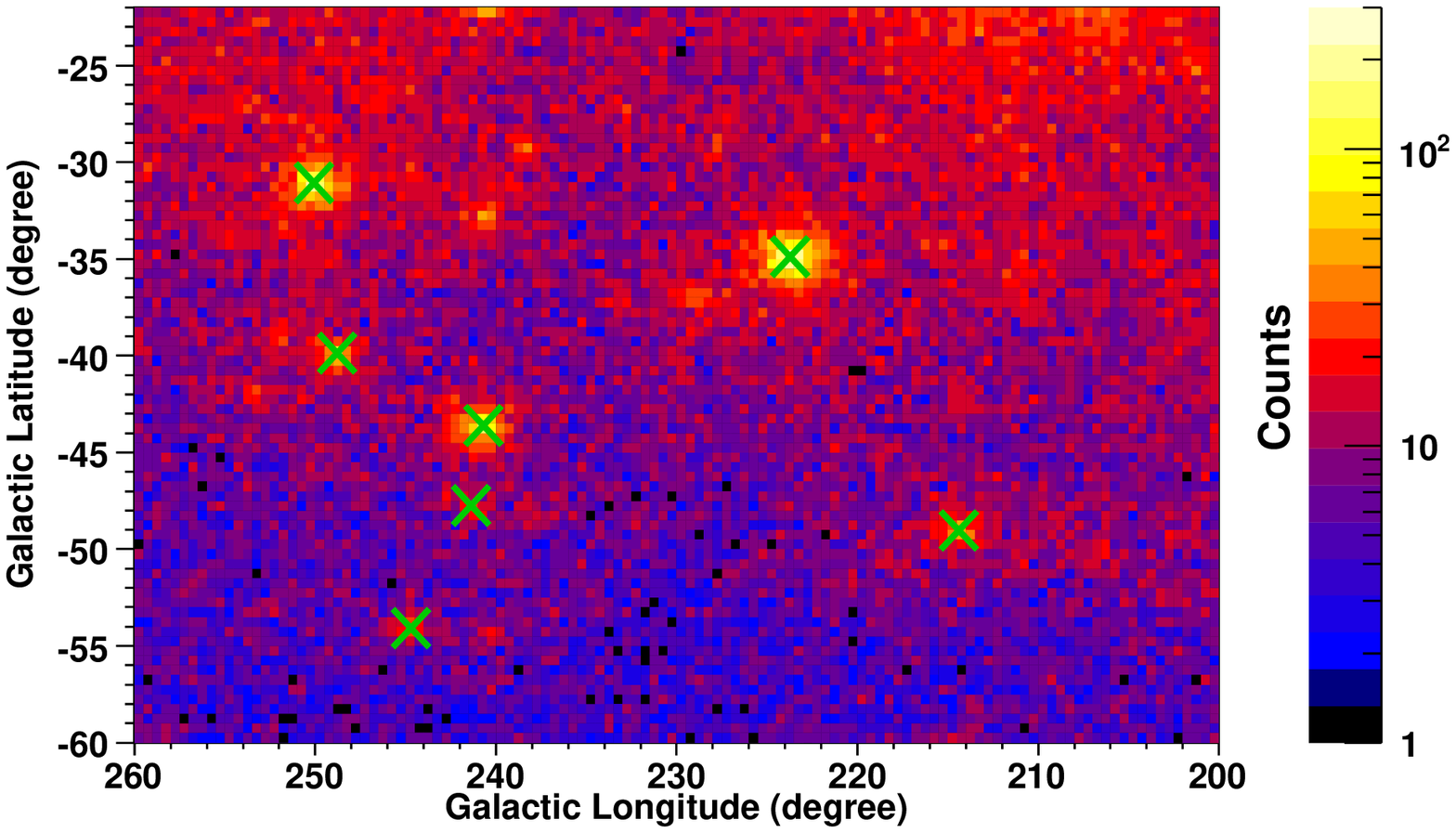}
\caption{\textit{Fermi} LAT $\gamma$-ray count maps ($E \ge 100~{\rm MeV}$) 
of regions we analyzed. Maps are in Cartesian projection
with $0.5\arcdeg \times 0.5\arcdeg$ binning.
The north and the south regions are shown in the upper panel 
and the lower panel, respectively.
In these regions there are 52 sources in the LAT 6 month 
source list,
of which 13 sources (indicated by green crosses) are
included in the LAT Bright Source List
\citep{Abdo2009b}.
}
\end{figure}

\begin{figure}
\plotone{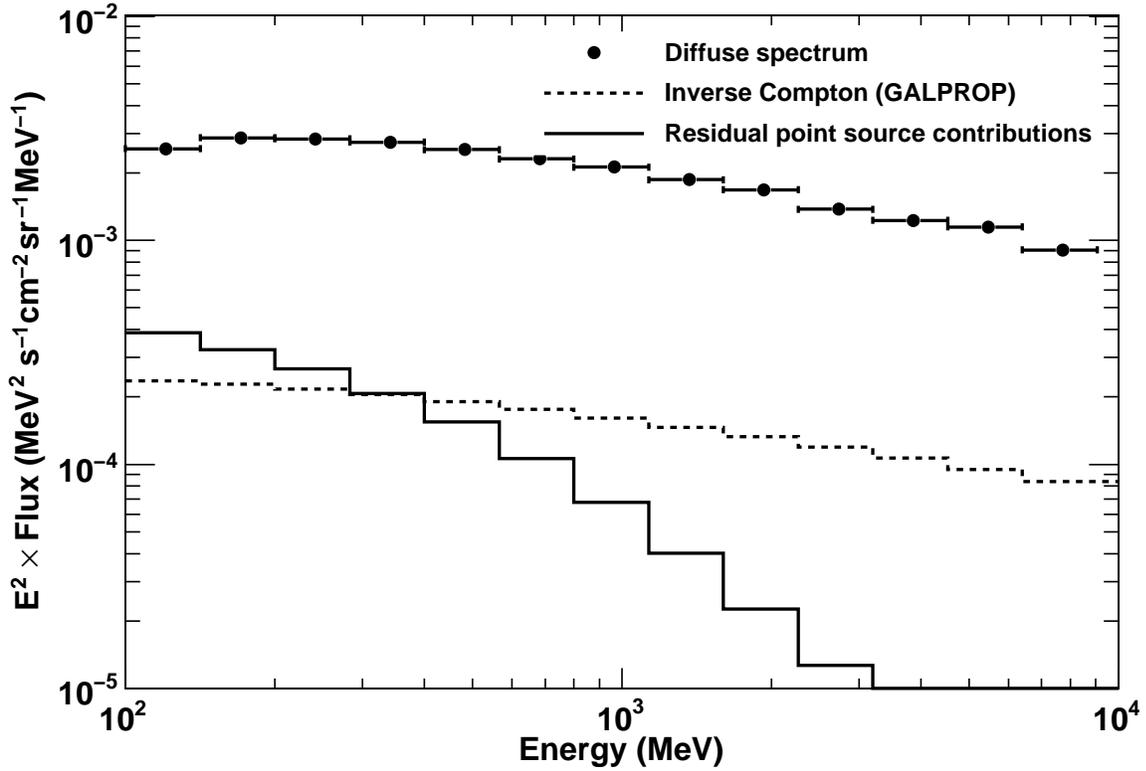}
\caption{
Diffuse $\gamma$-ray spectrum in regions analyzed 
($l$ from $200\arcdeg$ to $260\arcdeg$  and $\left| b \right|$
from $22\arcdeg$ to $60\arcdeg$ ) after masking point sources 
with circular regions of $1\arcdeg$ radius. 
IC emission predicted by GALPROP ({\bf 54\_5gXvarh7S}) 
and the residual point source contributions
estimated from spectral parameters given in the LAT 
6 month source list are shown 
by dotted and solid histograms, respectively.
The horizontal and vertical error bars indicate the energy ranges
and 1 $\sigma$ statistical errors, respectively.
}
\end{figure}

\begin{figure}
\plotone{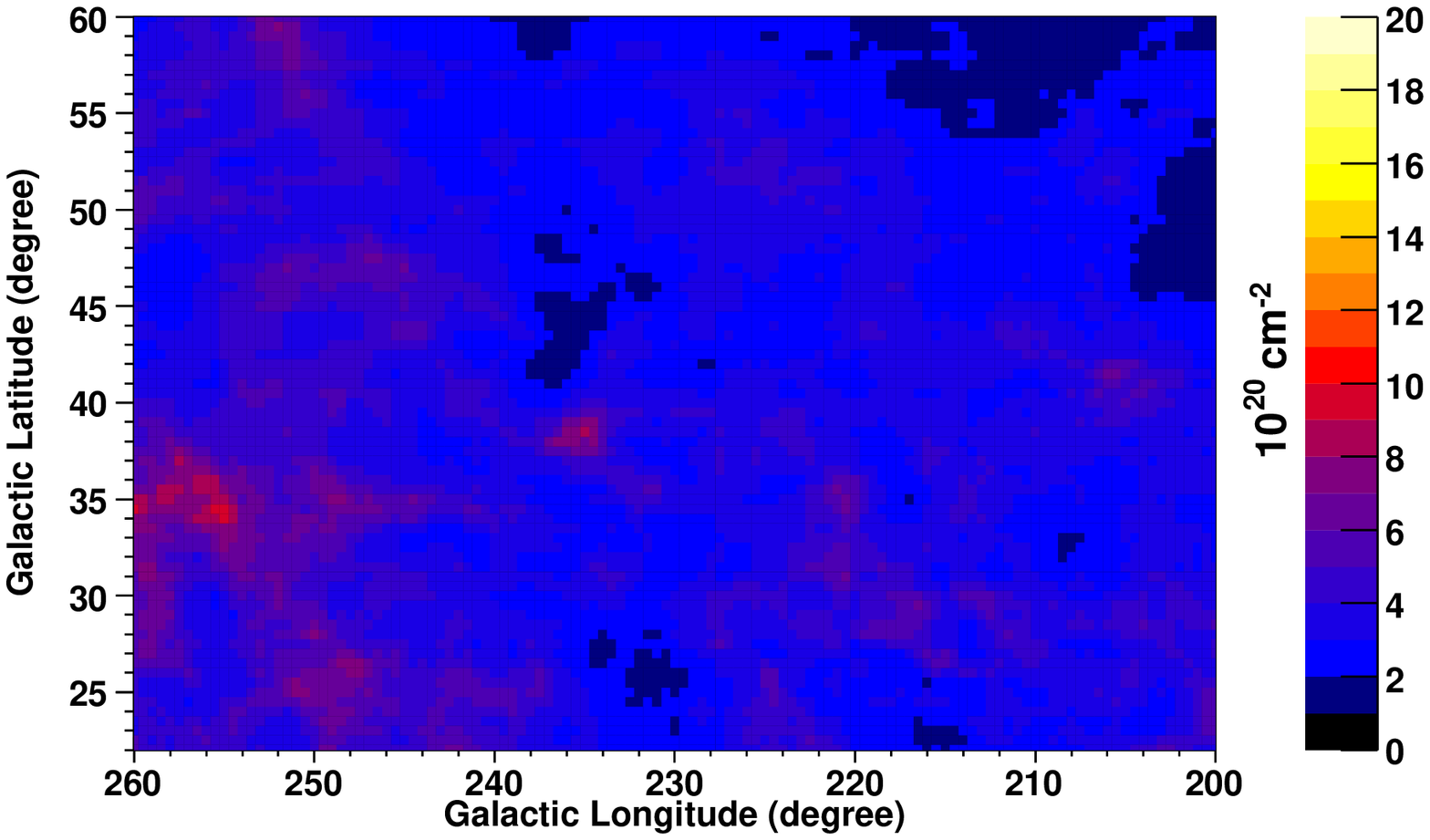}
\plotone{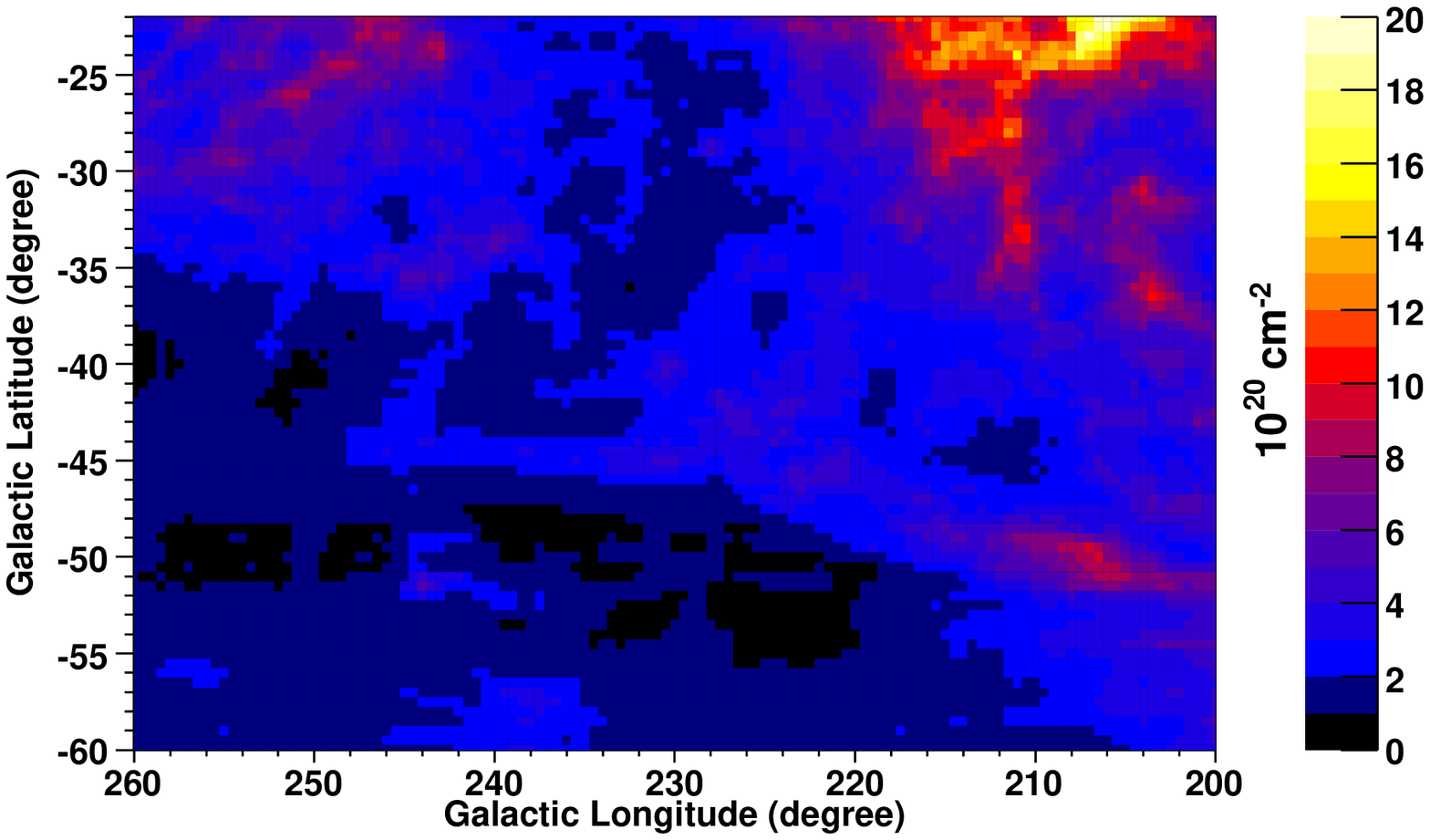}
\caption{
Column density maps of atomic hydrogen derived 
from the LAB survey under the assumption 
of a uniform spin temperature of 125~K. 
}
\end{figure}

\begin{figure}
%%\plotone{Result2_smallReg_1deg_dumpE2_fit_errFix.eps}
%%\plotone{Result2_smallReg_1deg_dumpE6_fit_errFix.eps}
%%\plotone{Result2_smallReg_1deg_dumpE10_fit_errFix.eps}
%%\plotone{Result2_smallReg_1deg_dumpE14_fit_errFix.eps}
\plottwo{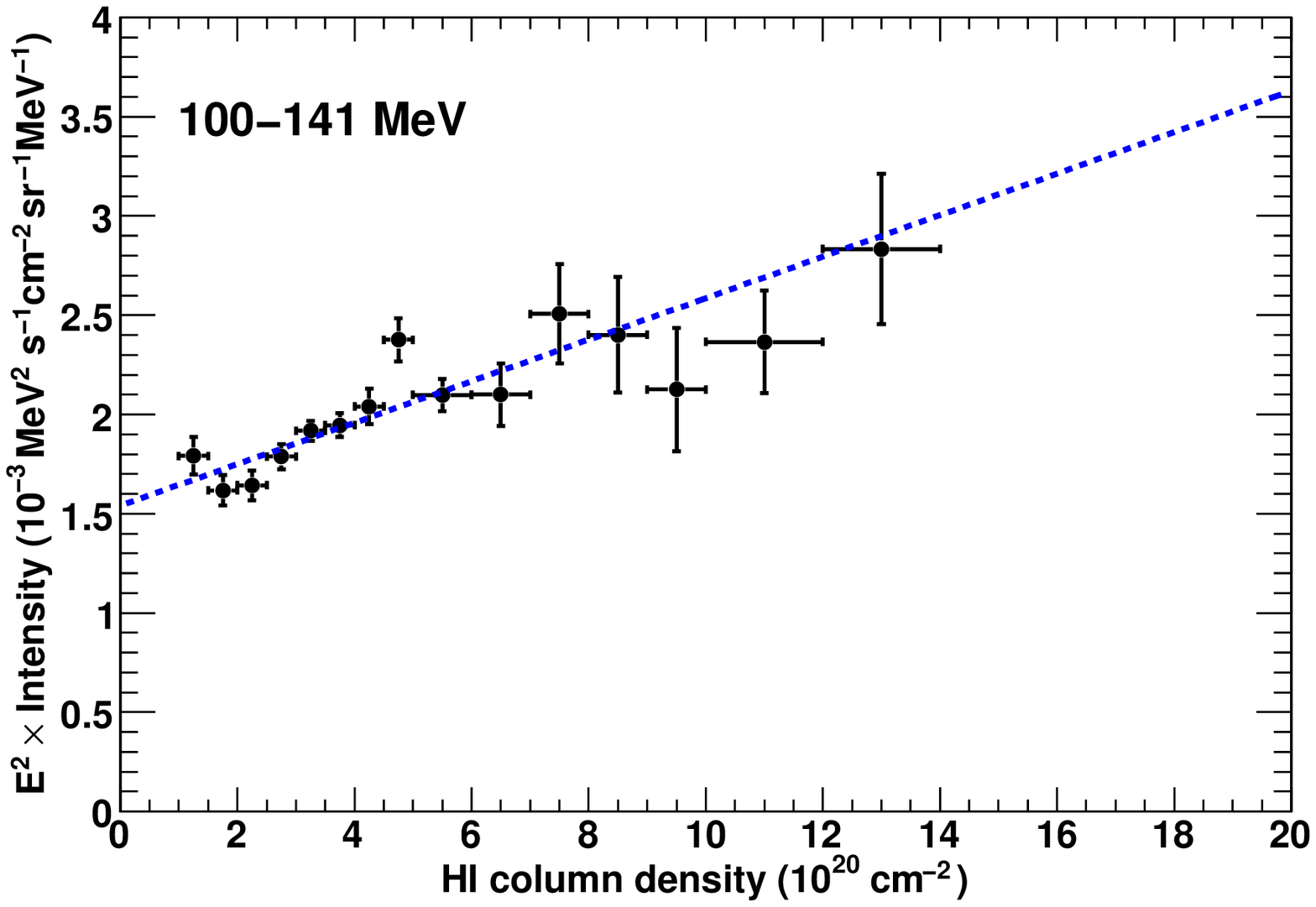}{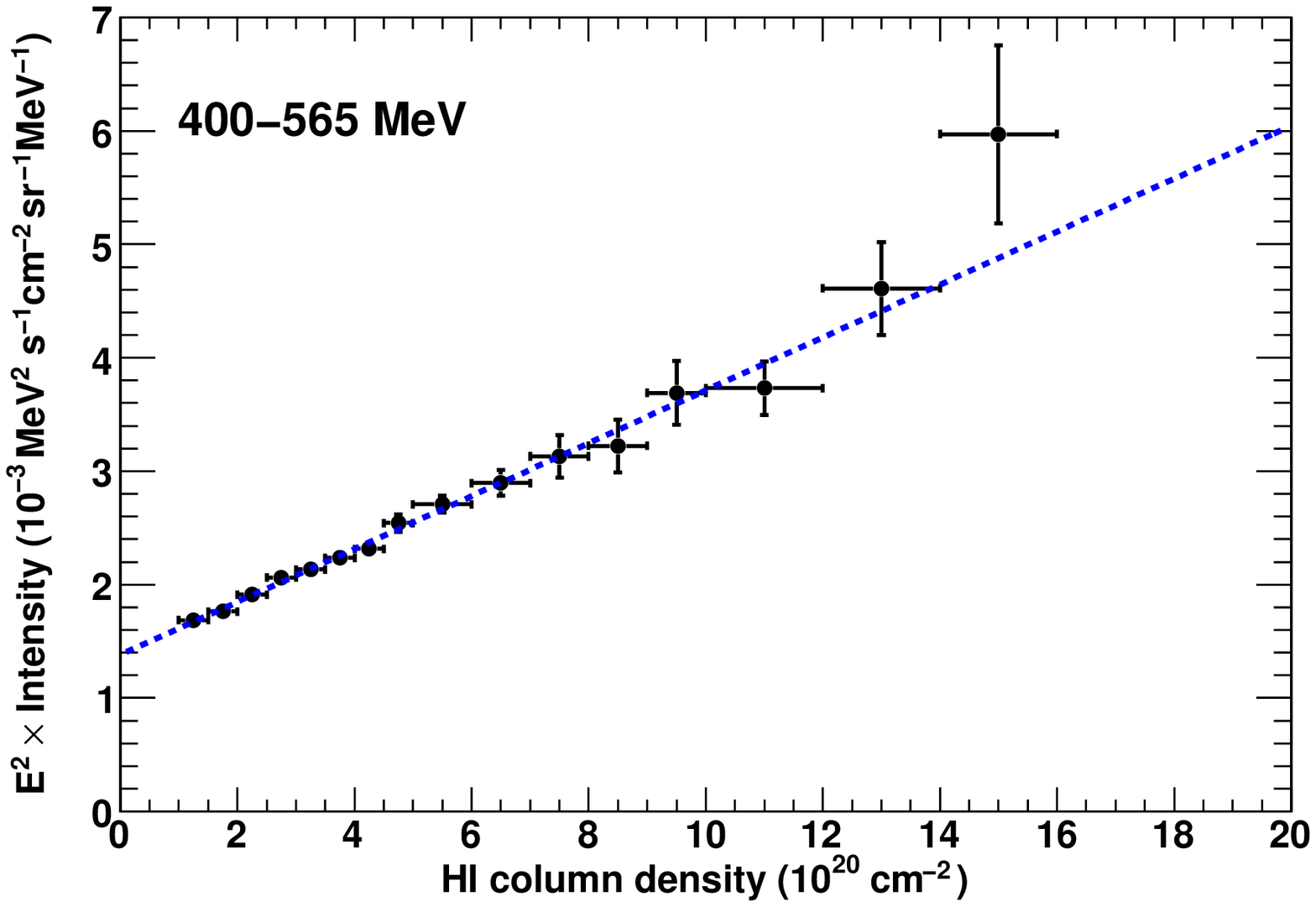} \\
\plottwo{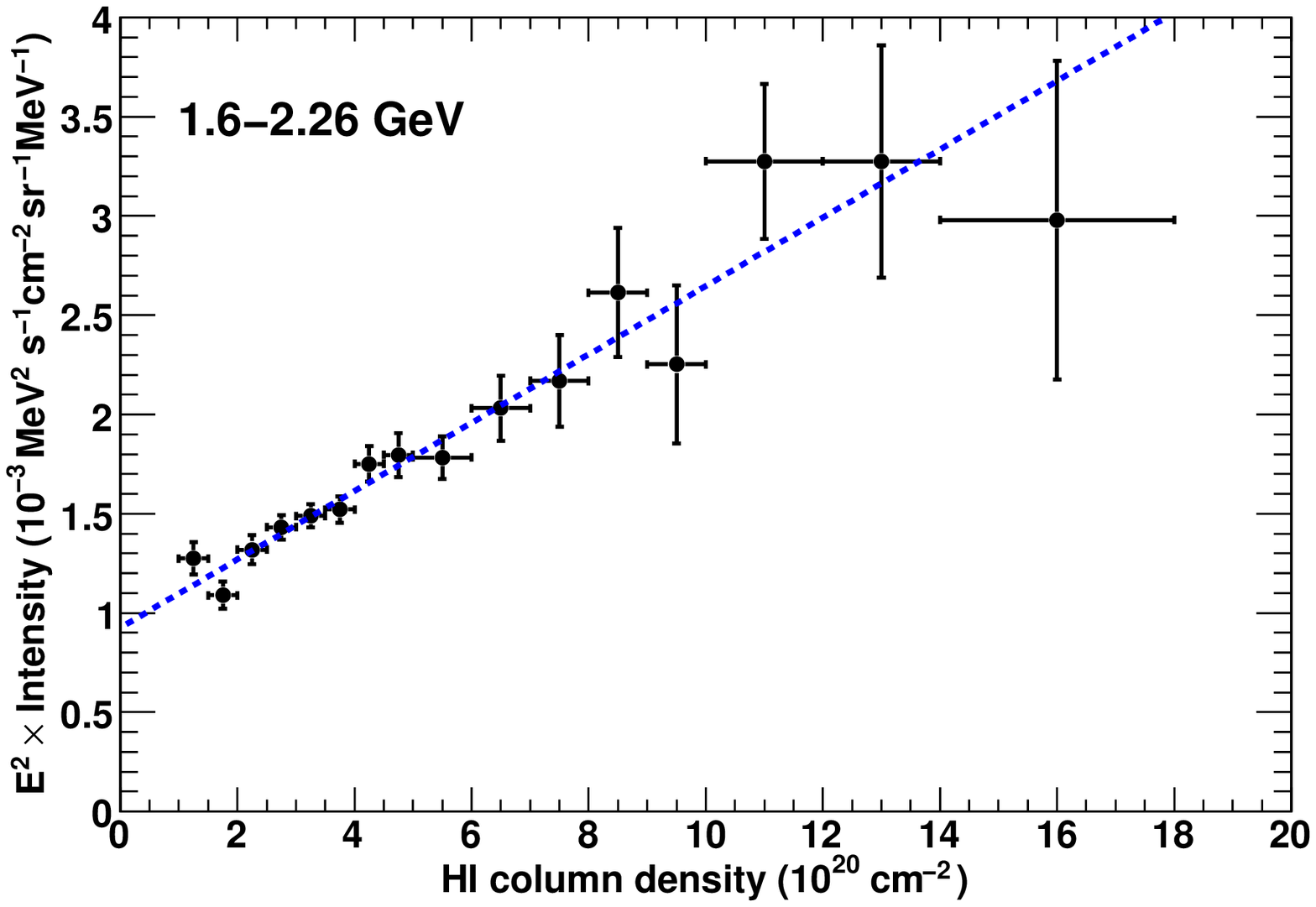}{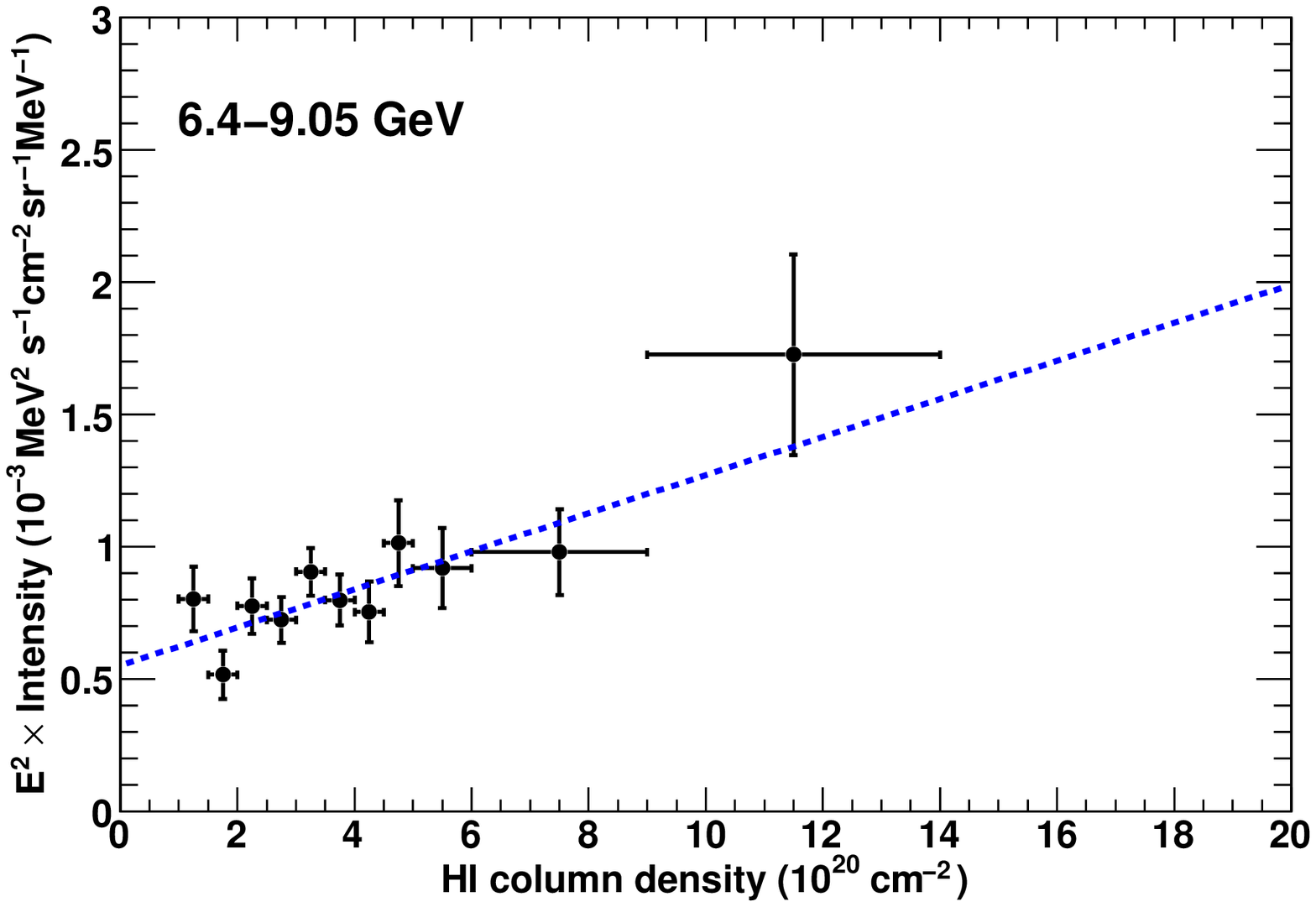}
\caption{
Correlation of the 
(IC and point-sources subtracted) $\gamma$-ray intensities and the 
\HI\ column densities in four representative bands. 
The map of $N$(\HI) (Figure 3) is convolved 
with the LAT PSF of the corresponding energy range.
The horizontal and vertical error bars indicate the 
ranges of the column density
and the 1 $\sigma$ statistical errors, respectively.
Data in high energy range (above 1~GeV) are rebinned
to have more than 10 $\gamma$-ray counts in each bin.
}
\end{figure}

\begin{figure}
\plotone{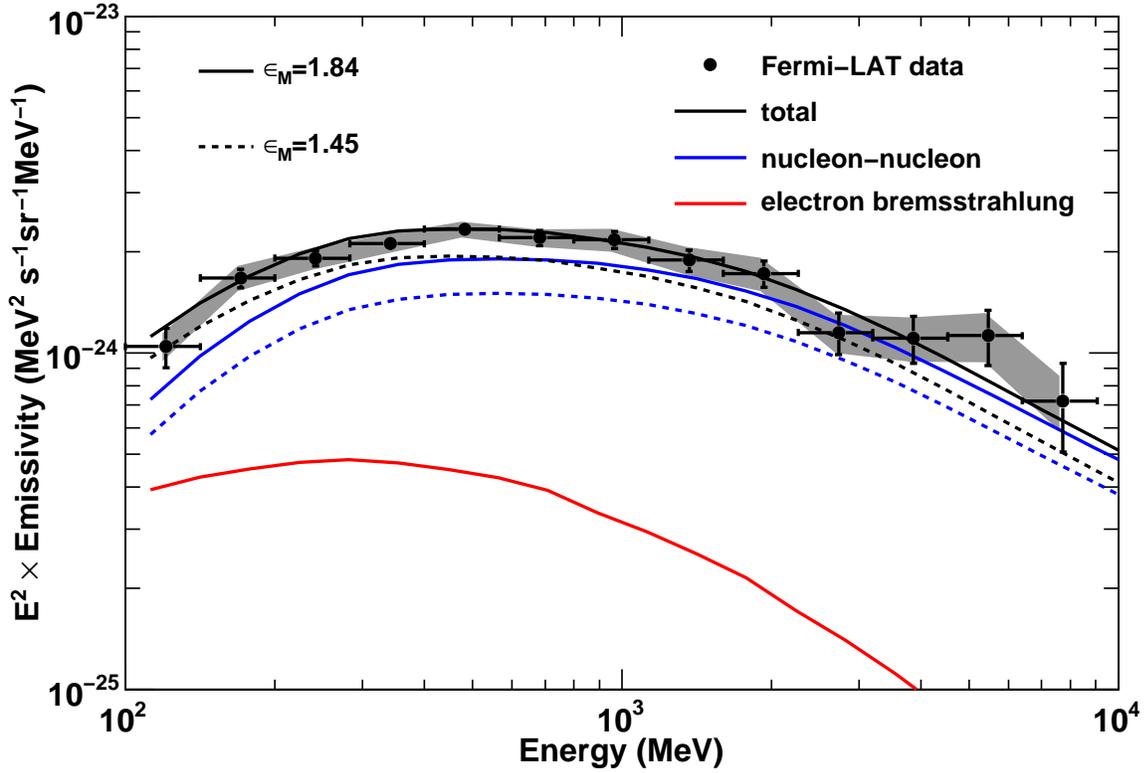}
\caption{
Differential $\gamma$-ray emissivity from the local atomic hydrogen gas 
compared with the calculated $\gamma$-ray production.
The horizontal and vertical error bars indicate
the energy ranges and 1 $\sigma$ statistical errors, respectively.
The assumed interstellar proton, electron and positron spectra are 
shown in Figure~6. Estimated systematic errors
of the LAT data are indicated by the shaded area.
A nucleus enhancement factor $\epsilon_{\rm M}$ of 1.84 is
assumed for the calculation of the $\gamma$-rays
from nucleon-nucleon interactions.
Dotted lines indicate the emissivities for the case of
$\epsilon_{\rm M}=1.45$, the lowest values
in the referenced literature.
}
\end{figure}

\begin{figure}
\plotone{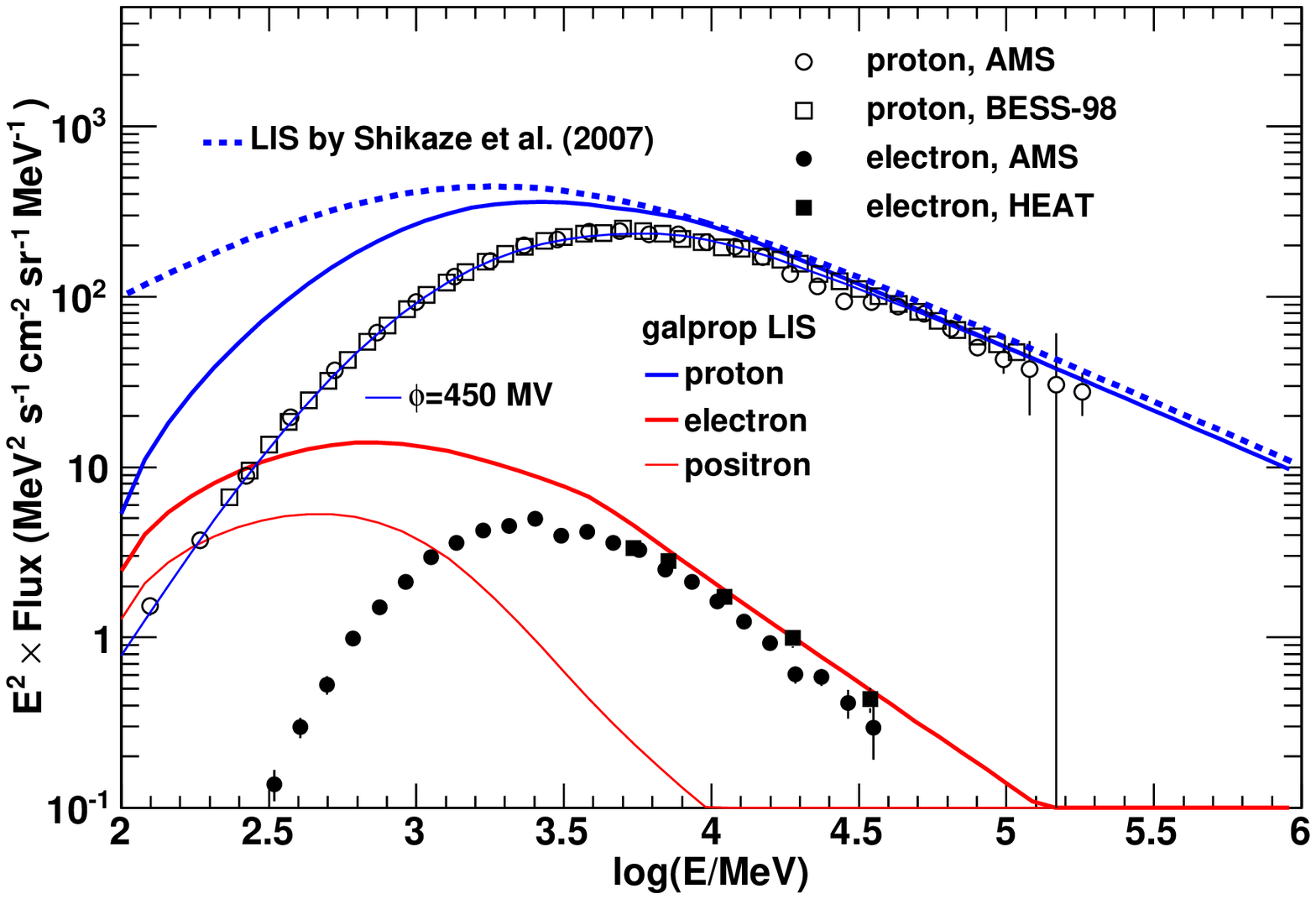}
\caption{
LIS of protons (thick blue line),
electrons (thick red line) and positrons (thin red line)
used to calculate 
the $\gamma$-ray emissivity model in Figure~5 
with a compilation of direct measurements at the Earth;
see \citet{Alcaraz2000a}, \citet{Sanuki2000} and
\citet{Shikaze2007} for proton data,
and \citet{Alcaraz2000b} and \citet{Barwick1998} for electron data. 
Modulated spectrum with ${\rm \phi = 450~MV}$ is given by
thin blue line. The proton LIS adopted by \citet{Shikaze2007}
is shown by dotted blue line.
}
\end{figure}

%% Tables should be submitted one per page, so put a \clearpage before

%% Two options are available to the author for producing tables:  the
%% deluxetable environment provided by the AASTeX package or the LaTeX
%% table environment.  Use of deluxetable is preferred.
%%

%% Three table samples follow, two marked up in the deluxetable environment,
%% one marked up as a LaTeX table.

%% In this first example, note that the \tabletypesize{}
%% command has been used to reduce the font size of the table.
%% We also use the \rotate command to rotate the table to
%% landscape orientation since it is very wide even at the
%% Note also that the \label command needs to be placed

%% This table also includes a table comment indicating that the full
%% version will be available in machine-readable format in the electronic

\clearpage

\begin{deluxetable}{cccccc}
\tabletypesize{\footnotesize}
\tablecaption{
A summary of fit parameters with 1 Sigma statistical errors. 
}
\tablehead{\colhead{Energy} & \colhead{Offset} & \colhead{Slope} &
\colhead{$\chi^{2}$/dof \tablenotemark{a}} & 
\colhead{Data/Model counts \tablenotemark{b}} \\
\colhead{GeV}& \colhead{${\rm 10^{-4}~MeV^{2}~s^{-1}~cm^{-2}~sr^{-1}~MeV^{-1}}$} & 
\colhead{${\rm 10^{-24}~MeV^{2}~s^{-1}~sr^{-1}~MeV^{-1}}$} & \colhead{} & \colhead{}}
\startdata
0.10--0.14 & $15.40 \pm 0.54$ & $1.04 \pm 0.14$ & 21.90/13 & 11799/11678 \\
0.14--0.20 & $17.10 \pm 0.40$ & $1.67 \pm 0.10$ & 18.12/14 & 27891/27738 \\
0.20--0.28 & $16.70 \pm 0.36$ & $1.91 \pm 0.09$ & 13.47/14 & 31718/31564 \\
0.28--0.40 & $15.83 \pm 0.36$ & $2.11 \pm 0.10$ & 16.92/14 & 28987/28850 \\
0.40--0.56 & $13.81 \pm 0.39$ & $2.33 \pm 0.10$ & \ 6.65/14 & 22718/22073 \\
0.56--0.80 & $12.57 \pm 0.41$ & $2.20 \pm 0.11$ & 16.17/15 & 16137/16063 \\
0.80--1.13 & $11.44 \pm 0.44$ & $2.17 \pm 0.12$ & 12.26/15 & 11421/11368 \\
1.13--1.60 & $10.23 \pm 0.49$ & $1.88 \pm 0.13$ & \ 9.06/14 &  7364/7327  \\
1.60--2.26 & $\ 9.25 \pm 0.54$ & $1.72 \pm 0.15$ & 13.16/14 &  4765/4733  \\
2.26--3.20 & $\ 8.44 \pm 0.58$ & $1.15 \pm 0.16$ & 10.56/12 &  2764/2747  \\
3.20--4.53 & $\ 7.12 \pm 0.64$ & $1.10 \pm 0.17$ & 15.07/11 &  1733/1712  \\
4.53--6.40 & $\ 6.44 \pm 0.75$ & $1.12 \pm 0.21$ & 11.35/11 &  1158/1144  \\
6.40--9.05 & $\ 5.51 \pm 0.77$ & $0.71 \pm 0.21$ & \ 9.92/9  &   678/664   \\
\enddata
\tablenotetext{a}{
Degree of freedom}
\tablenotetext{b}{
Data and model total counts
after masking point sources with
circular regions of $1\arcdeg$ radius.
We believe that the small ($\le 1$~\%) excesses of the data counts 
over the model counts are due to unresolved point sources 
or interstellar matter
not traced by 21~cm line surveys.}
\end{deluxetable}

\clearpage

\end{document}